\documentclass[10pt]{article}
\usepackage{epsfig,graphicx}
\usepackage{times,ch2005}

\def\beq{\begin{equation}}
\def\eeq#1{\label{#1}\end{equation}}
\def\eeqn{\end{equation}}

\def\beqa{\begin{eqnarray}}
\def\eeqa#1{\label{#1}\end{eqnarray}}
\def\eeqan{\end{eqnarray}}

\let\bar=\overbar

\def\Dslash{\not{\hbox{\kern-4pt $D$}}}
\def\dslash{\not{\hbox{\kern-2pt $\del$}}}

\newcommand{\tev}{\ensuremath{\mathrm{\,Te\kern -0.1em V}}\xspace}
\newcommand{\gev}{\ensuremath{\mathrm{\,Ge\kern -0.1em V}}\xspace}
\newcommand{\mev}{\ensuremath{\mathrm{\,Me\kern -0.1em V}}\xspace}
\newcommand{\kev}{\ensuremath{\mathrm{\,ke\kern -0.1em V}}\xspace}
\newcommand{\ev}{\ensuremath{\mathrm{\,e\kern -0.1em V}}\xspace}
\newcommand{\gevc}{\ensuremath{{\mathrm{\,Ge\kern -0.1em V\!/}c}}\xspace}
\newcommand{\mevc}{\ensuremath{{\mathrm{\,Me\kern -0.1em V\!/}c}}\xspace}
\newcommand{\gevcc}{\ensuremath{{\mathrm{\,Ge\kern -0.1em V\!/}c^2}}\xspace}
\newcommand{\mevcc}{\ensuremath{{\mathrm{\,Me\kern -0.1em V\!/}c^2}}\xspace}

\def\mus  {\ensuremath{\rm \,\mus}\xspace}

\def\mus        {\ensuremath{\,\mu{\rm s}}\xspace}    

\begin{document}


\Title{Flux Measurement at 110 GeV\\\vspace{0.2\baselineskip} of the Blazar Mrk~501 with CELESTE}
\bigskip


%
\label{Brion}

%
\author{Elisabeth Brion\index{Brion, E.} for the CELESTE Collaboration}

%
\address{Centre d'\'etudes nucl\'eaires de Bordeaux-Gradignan (CENBG)\\
Chemin du Solarium~-- Le Haut-Vigneau~-- BP 120\\
33175 Gradignan Cedex, France\\
}

\makeauthor\abstracts{The new analysis variable $\xi$, shown to be powerful on the data taken with the final configuration of CELESTE, has been applied to data taken with previous detector con\-fi\-gu\-ra\-tions. First, the analysis is validated on Crab observations, and then the cuts for the blazar Mrk~501 are optimized using Mrk~421 data since the sources have similar declinations. Data from Mrk~501 was recorded in 2000 and 2001. The old analysis gave a $2.5~\sigma$ excess. We obtain an excess of $2.9~\sigma$ during this time and of $4.9~\sigma$ during May and June~2000 that we interpret as a $\gamma$-ray signal from Mrk~501, for which we calculate a flux of $(6.9\pm2.2)\times10^{-7}~\mathrm{photons\,m^{-2}\,s^{-1}}$. An upper limit from the other data with no signal is determined.}

\section{Introduction}

CELESTE (Cherenkov Low Energy Sampling and Timing Experiment) was a Che\-ren\-kov experiment using the heliostats of the former \'Electricit\'e de France solar plant in the French Pyrenees at the Th\'emis site. It detected Cherenkov light from showers produced in the atmosphere by cosmic rays and the $\gamma$-rays coming from high energy astrophysical sources. The light is reflected to secondary optics and photomultipliers installed at the top of the tower. Finally it is sampled to be analysed~\cite{Pare}.

Two states of the experiment have to be distinguished to classify CELESTE data. During the first one (between September 1999 and June 2001), 40~heliostats were used with two types of pointing (single pointing: all heliostats at $11~\mathrm{km}$, double pointing: half at $11~\mathrm{km}$, half at $25~\mathrm{km}$), and during the second one (between September 2001 and June 2004), 53~heliostats pointing at $11~\mathrm{km}$ were used (of which 12 veto heliostats aimed wide for proton rejection). An analysis improvement was made using 53~heliostat data for the Crab Nebula, a bright and stable source, the new analysis variable provided a sensitivity of $5.7~\sigma/\sqrt{\mathrm{h}}$ on $4.7~\mathrm{h}$ data, whereas the old data analysis gave $2.0$ and $3.4~\sigma/\sqrt{\mathrm{h}}$ for single and double pointing~\cite{MdN}. The new analysis has been tested on 40~heliostat Crab data and gives better sensitivity as will be shown.

CELESTE has taken 40~heliostat data on the blazar Mrk~501. The old analysis gave $2.5~\sigma$~\cite{LeGallou}. We present here the results obtained with the new analysis after optimizing cuts on the blazar Mrk~421 which has nearly the same declination as Mrk~501.

\section{New Analysis applied to 40~Heliostat Crab Data}

After data selection based on stability criteria, a software trigger is applied to the data in order to remove trigger bias. The data can then be analysed with the new variable $\xi$ which is a normalised sum of the Cherenkov pulses from each heliostat, optimised by finding the shower position that gives the narrowest sum (i.e. the most gamma-like)~\cite{Manseri, Bruel}. $\xi$ has been developed on Crab data taken with 53~heliostats. The 40~heliostats data sets have been analysed with new cuts on $\xi$, determined from simulations and data. The study confirmed the dependence of $\xi$ on configuration (heliostat number, trigger) and pointing (hour angle). Different cuts were chosen depending on these parameters, yielding a sensitivity of $4.0~\sigma/\sqrt{\mathrm{h}}$ and $5.2~\sigma/\sqrt{\mathrm{h}}$ for single and double pointing with 40~heliostats.

\begin{figure}[!ht]
   \centering\includegraphics[bb=14 14 609 857, clip=true, viewport=80 18 558 710, angle=270, width=0.49\textwidth]{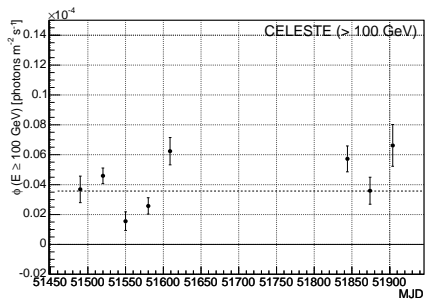}
   \caption{Crab nebula lightcurve seen by CELESTE above $100~\mathrm{GeV}$ with statistical error bars (monthly averages centred on New Moon). The dashed line represents the mean integral flux for all the data: $\bar\phi(E\,{\geq}\,100~\mathrm{GeV})=(3.6\pm0.2)\times10^{-6}~\mathrm{photons\,m^{-2}\,s^{-1}}$). $\mathrm{MJD}$ (Modified Julian Date) $51450$ and $51900$ correspond to $09/29/1999$ and $12/22/2000$.}
   \label{fig:brion-fig1}
\end{figure}

The acceptances of the detector were calculated to determine a flux. They depend on many parameters: atmosphere, optical simulation (heliostats, secondary mirrors), electronic simulation, and pointing (configuration, declination, hour angle). The si\-mu\-la\-tions have been improved to better fit the data, taking into account variations of the trigger rates between different data sets (two different atmospheres) and degradation of the optical chain of the detector. The lightcurve for the Crab nebula was determined for the 40~heliostat data with a spectral hypothesis of $E^{\alpha+\beta\log E}$ with $E$ in $\mathrm{TeV}$, $\alpha=-2.74$ and $\beta=-0.50$~\cite{MdN}. It is shown in figure~\ref{fig:brion-fig1}, where the calculated flux is stable, within the statistical error bars.

\section{Mrk~501 Flux Determination after Cut Optimisation using Mrk~421 Observations}

The blazar Mrk~421 has the double advantage of having about the same declination as Mrk~501 and of being well observed with CELESTE. So cuts were optimized on Mrk~421 data. (In fact, both the simulation and the data lead to very nearby the same cuts.) We obtained a final significance for all Mrk~421 data of $29.4~\sigma$ during $38.9~\mathrm{h}$. The integral flux is $\bar\phi(E\,{\geq}\,100~\mathrm{GeV})=(2.3\pm0.1)\times10^{-6}~\mathrm{photons\,m^{-2}\,s^{-1}}$ for an $E^{-2}$ spectral hypothesis.

\begin{figure}[!ht]
\begin{minipage}[c]{0.49\textwidth}
\begin{center}
   \includegraphics[bb=14 14 609 857, clip=true, viewport=8 410 506 730, width=\textwidth]{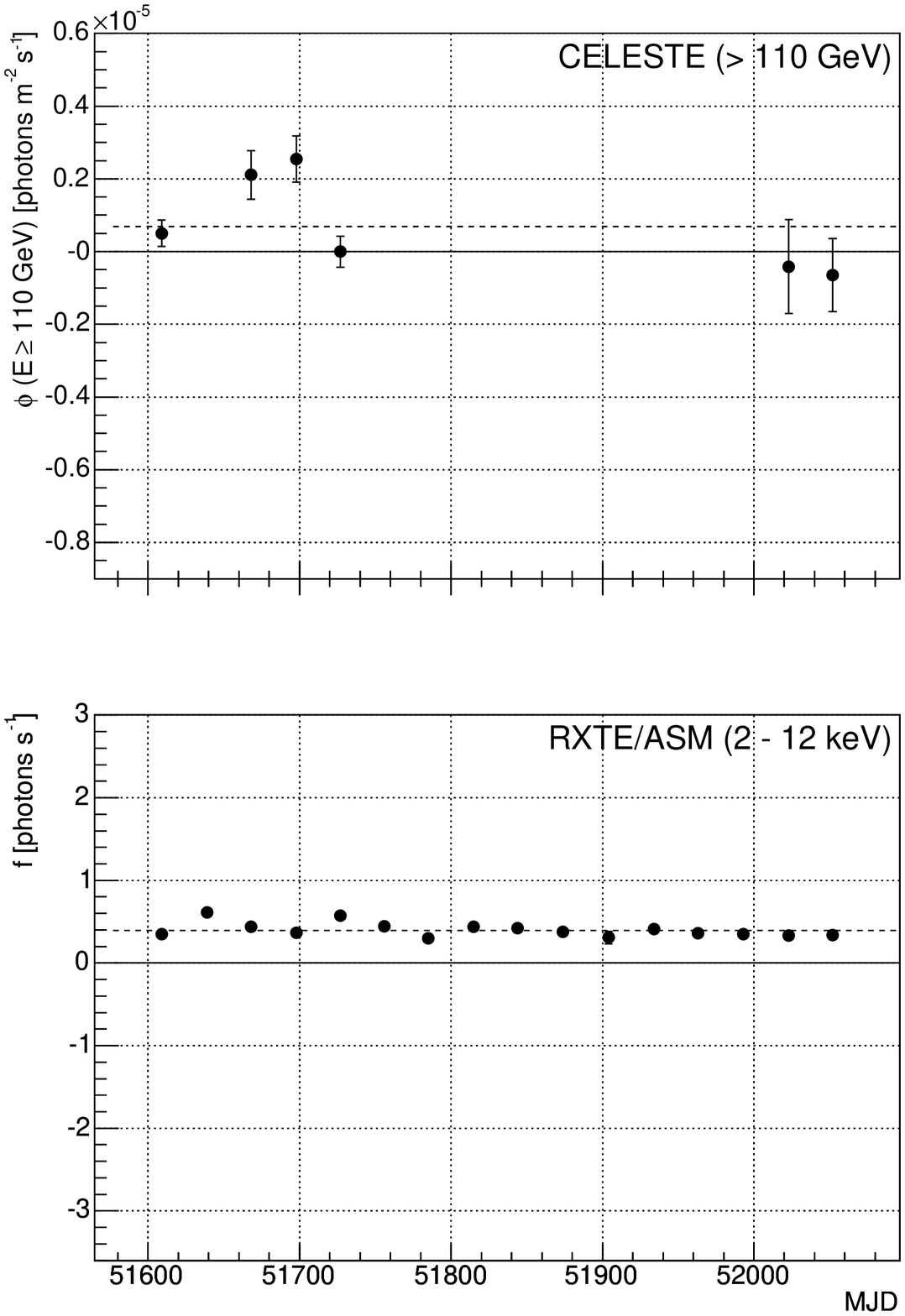}
   \includegraphics[bb=14 14 609 857, clip=true, viewport=8 16 506 355, width=\textwidth]{brion-fig2.ps}
\end{center}
\end{minipage}
\begin{minipage}[c]{0.49\textwidth}
   \centering\includegraphics[width=\textwidth]{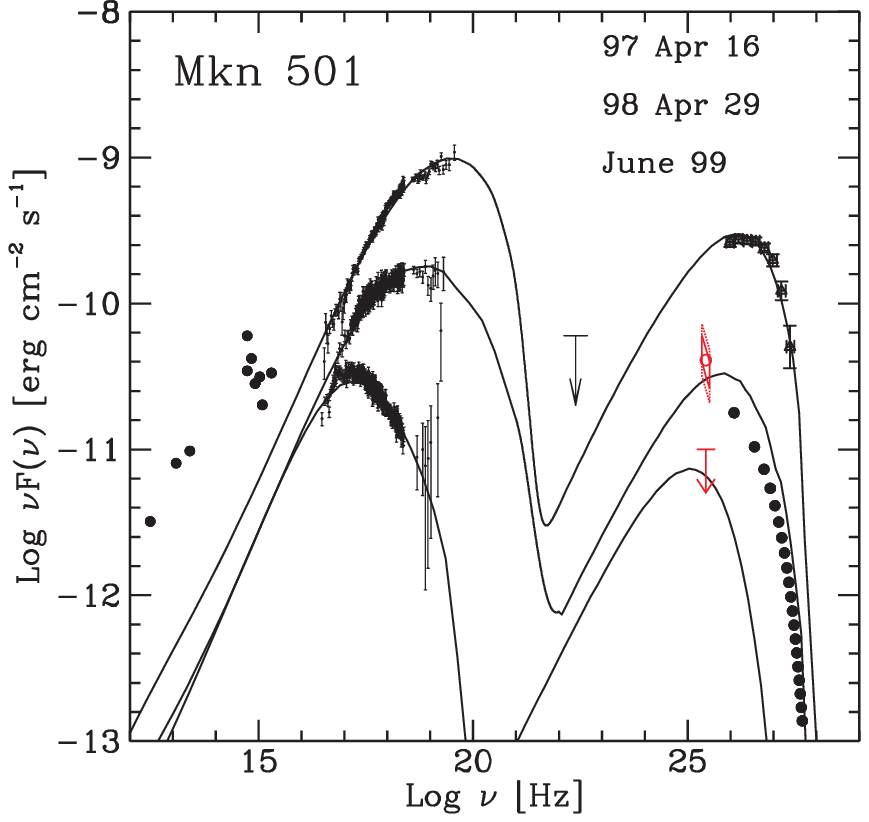}
\end{minipage}
   \caption{Left: Mrk~501 lightcurves (monthly averages centred on New Moon) seen by CELESTE ($\bar\phi(E\,{\geq}\,110~\mathrm{GeV})=(6.9\pm2.2)\times10^{-7}~\mathrm{photons\,m^{-2}\,s^{-1}}$, top) and RXTE/ASM in X-rays ($\bar{f}=0.40\pm0.01~\mathrm{photons\,s^{-1}}$, bottom) with statistical error bars. $\mathrm{MJD}$ $51600$ and $52000$ correspond to $02/26/2000$ and $04/01/2001$. Right: overall spectral energy distribution of Mrk~501 of 04/16/1997, 04/29/1998 and June 1999~\cite{Tavecchio}. Red circle (with statistical and systematic error bars) and arrow are the CELESTE flux and upper limit determined in this work at $2.6\times10^{25}~\mathrm{Hz}$.}
   \label{fig:brion-fig2}
\end{figure}

The number of events and significance for Mrk~501 after the same cuts are shown in table~\ref{tab:brion-Mrk501}. A $2.9~\sigma$ excess is obtained for the 2000-2001 period. The lightcurve of Mrk~501 (figure~\ref{fig:brion-fig2} left top), determined for an $E^{-2}$ spectral hypothesis, presents an excess of activity during May and June~2000 (between MJD 51660 and 51700). For this period, the significance of the excess is $4.9~\sigma$ which is interpreted as a $\gamma$-ray signal from Mrk~501, since the blazar is already well-known in low- and high-energy $\gamma$-rays. Thus, a differential flux is calculated for this period and an upper limit is determined for the rest of the data which contains no signal.

\begin{table}[htbp]
\begin{center}
\resizebox{\hsize}{!}{
\begin{tabular}{|l|c|c|c|c|c|}
\hline
Cut & \multicolumn{3}{|c|}{Number of events} & Significance & Signal to noise ratio\\
& $N_\mathrm{\textit{\scriptsize{ON}}}$ & $N_\mathrm{\textit{\scriptsize{OFF}}}$ & $N_\mathrm{\textit{\scriptsize{ON}}}-N_\mathrm{\textit{\scriptsize{OFF}}}$ & $N_\sigma$ & $\displaystyle\frac{N_\mathrm{\textit{\scriptsize{ON}}}-N_\mathrm{\textit{\scriptsize{OFF}}}}{N_\mathrm{\textit{\scriptsize{OFF}}}}~\mathrm{[\%]}$\\
\hline \hline
Raw trigger & $815\,924$ & $813\,641$ & $2\,284$ & $1.5$ & $0.3$\\
Software trigger & $472\,450$ & $469\,569$ & $2\,881$ & $2.5$ & $0.6$\\
Final cuts & $10\,565$ & $10\,096$ & $469$ & $2.9$ & $4.6$\\
\hline
\end{tabular}
}
\caption{Number of events for the blazar Mrk~501 after cuts for the final data set ($11.5~\mathrm{h}$).}
\label{tab:brion-Mrk501}
\end{center}
\end{table}

\section{Discussion}

The differential flux and the upper limit, obtained for a counting maximum at $110~\mathrm{GeV}$, are drawn on figure~\ref{fig:brion-fig2} right. The different states of activity of Mrk~501 in $\gamma$-rays are not correlated to variations in X-rays: in figure~\ref{fig:brion-fig2} left bottom, the ASM flux is $40\pm10~\mathrm{\%}$ higher in May and June~2000 than the low baseline of the rest of the studied period (figure~\ref{fig:brion-fig2} right shows a $\times20$ and $\times100$ increases during historical flares. These observations suggest that for May and June~2000, the emission population was different between synchrotron emission in X-rays and $\gamma$-ray emission. If the $\gamma$-ray emission is due to an inverse Compton process, it can not be explained with a SSC (Synchrotron Self-Compton) model for this particular case. Further observations of Mrk~501 with new generation detectors (\textit{GLAST}, HESS) and simultaneous to other wavelength observations will help to understand the emission processes of Mrk~501.

\label{Brion}
 
\end{document}